\documentclass[conference]{IEEEtran}
\IEEEoverridecommandlockouts
\usepackage{cite}
\usepackage{amsmath,amssymb,amsfonts}
\usepackage{algorithmic}
\usepackage{graphicx}
\usepackage{textcomp}
\usepackage{xcolor}
\usepackage{hyperref}
\usepackage{listings}
\usepackage{float}
\usepackage{algorithmic}
\usepackage[linesnumbered,ruled,vlined]{algorithm2e}
\usepackage{tikz}
\usepackage{pgfplots}
\usepackage{balance}
\pgfplotsset{compat=1.18}

\def\BibTeX{{\rm B\kern-.05em{\sc i\kern-.025em b}\kern-.08em
    T\kern-.1667em\lower.7ex\hbox{E}\kern-.125emX}}

\newcommand\YAMLcolonstyle{\color{red}\mdseries}
\newcommand\YAMLkeystyle{\color{black}\ttfamily\footnotesize}
\newcommand\YAMLvaluestyle{\color{blue}\mdseries}

\makeatletter

\newcommand\language@yaml{yaml}

\expandafter\expandafter\expandafter\lstdefinelanguage
\expandafter{\language@yaml}
{
  keywords={true1,false1,null1,y1,n1},
  keywordstyle=\color{darkgray}\bfseries,
  basicstyle=\YAMLkeystyle,                                 
  sensitive=false,
  comment=[l]{\#},
  morecomment=[s]{/*}{*/},
  commentstyle=\color{purple}\ttfamily,
  stringstyle=\YAMLvaluestyle\ttfamily,
  moredelim=[l][\color{orange}]{\&},
  moredelim=[l][\color{magenta}]{*},
  moredelim=**[il][\YAMLcolonstyle{:}\YAMLvaluestyle]{:},   
  morestring=[b]',
  morestring=[b]",
  literate =    {---}{{\ProcessThreeDashes}}3
                {>}{{\textcolor{red}\textgreater}}1
                {|}{{\textcolor{red}\textbar}}1
                {\ -\ }{{\mdseries\ -\ }}3,
  xleftmargin=2em,
}

\lst@AddToHook{EveryLine}{\ifx\lst@language\language@yaml\YAMLkeystyle\fi}
\makeatother

\begin{document}

\title{Simplifying HPC resource selection: A tool for optimizing execution time and cost on Azure}

\author{\IEEEauthorblockN{Marco A. S. Netto}
\IEEEauthorblockA{\textit{Microsoft}\\
Redmond, USA\\
marconetto@microsoft.com}
\and
\IEEEauthorblockN{Wolfgang De Savador}
\IEEEauthorblockA{\textit{Microsoft}\\
Rome, Italy\\
wolfgang.desalvador@microsoft.com}
\and
\IEEEauthorblockN{Davide Vanzo}
\IEEEauthorblockA{\textit{Microsoft}\\
Grapevine, USA \\
davide.vanzo@microsoft.com}
}

\maketitle

\begin{abstract}

Azure Cloud offers a wide range of resources for running HPC workloads,
requiring users to configure their deployment by selecting VM types, number
of VMs, and processes per VM. Suboptimal decisions may lead to longer
execution times or additional costs for the user. We are developing an
open-source tool to assist users in making these decisions by considering
application input parameters, as they influence resource consumption. The
tool automates the time-consuming process of setting up the cloud
environment, executing the benchmarking runs, handling output, and
providing users with resource selection recommendations as high level
insights on run times and costs across different VM types and number of
VMs. In this work, we present initial results and insights on reducing the
number of cloud executions needed to provide such guidance, leveraging data
analytics and optimization techniques with two well-known HPC applications:
OpenFOAM and LAMMPS.

\end{abstract}

\begin{IEEEkeywords}
HPC, Cloud, Resource Selection, Benchmarking, OpenFOAM, LAMMPS
\end{IEEEkeywords}

\section{Cloud Resource Selection}

Access to HPC resources through the cloud is becoming increasingly relevant in
different industries, research, and engineering institutions.
While delivering access to a flexible capacity on-demand, cloud imposes
significant pressure on performance tuning and cost-effectiveness of workload execution. With resources available on a flexible, per-minute billing basis, it is essential to configure job execution to minimize both wall-clock time and cost, finding the optimal balance.

In the cloud, if scalability remains near ideal, reducing resources will not lower job costs. Sometimes, scalability can be super-linear, where more nodes and cores improve performance due to memory bandwidth amplification, especially for specific cases like CFD or Explicit FEA.  In these scenarios, using more nodes may be cheaper if the licensing model permits, allowing faster results.

Latest architectures, for example AMD chiplet-based CPUs like Naples/Rome/Milan/Genoa, bring a huge core density possible on a single socket. On memory-bound workloads, selecting the optimal CPU per node is crucial for maximizing memory bandwidth per core. Moreover, technologies like AMD L3 Cache provide significant memory bandwidth amplification to be taken into consideration.

An optimal selection of the number of cores and process placement is always critical for the best cost/performance optimized execution~\cite{betting2023oikonomos,brunetta2019selecting,lamar2023evaluating,samuel2020a2cloud}. However, achieving that selection is not trivial for users, especially those who are not IT experts.

\section{HPCAdvisor and Scenario Optimization}

In this poster, we describe an open-source tool\footnote{\url{https://azure.github.io/hpcadvisor/}}  to help users select Cloud resources considering VM types, number of VMs, processes per VM, and application input parameters. The tool performs the end-to-end cycle of setting up a cloud environment, running scenarios, organizing the output, generating plots, and providing the advice as a Pareto front with execution time and costs as objectives.

The tool currently assumes no or little data available from previous
executions---it relies on data analytics and optimization techniques
to reduce the number of scenarios to be used in the cloud. As we embrace cases
where substantial data is available, we will
consider machine learning techniques too. We created a predictor that
reduces the scenarios by exploring two cases: (i) same application input but different VM types; and (ii) same VM type and different application input.

The first case leverages known data points---original VM type for
the tested number of nodes, and one or two data points for the target VM type; all with the same application input.
We predict the execution times by scaling the data points of the original VM type with
an optimal scaling factor, which is calculated using an
objective function that penalizes deviations between known data points of
the original VM type and predicted times, obtained via linear interpolation
across the segments of the curve for different number of nodes. The function
then uses the Broyden-Fletcher-Goldfarb-Shanno (BFGS) iterative method to
optimize the scaling factor for the best fit, and finally generates a new
curve. For the second case, we use application input (i.e. number of atoms for LAMMPS or number of cells for OpenFOAM) as
multiplication factor for generating the new data points, as such input has a direct impact on execution time.

\section{Experiments and Findings}

\begin{figure}[!h]
        \centering
       \includegraphics[width=1.0\columnwidth]{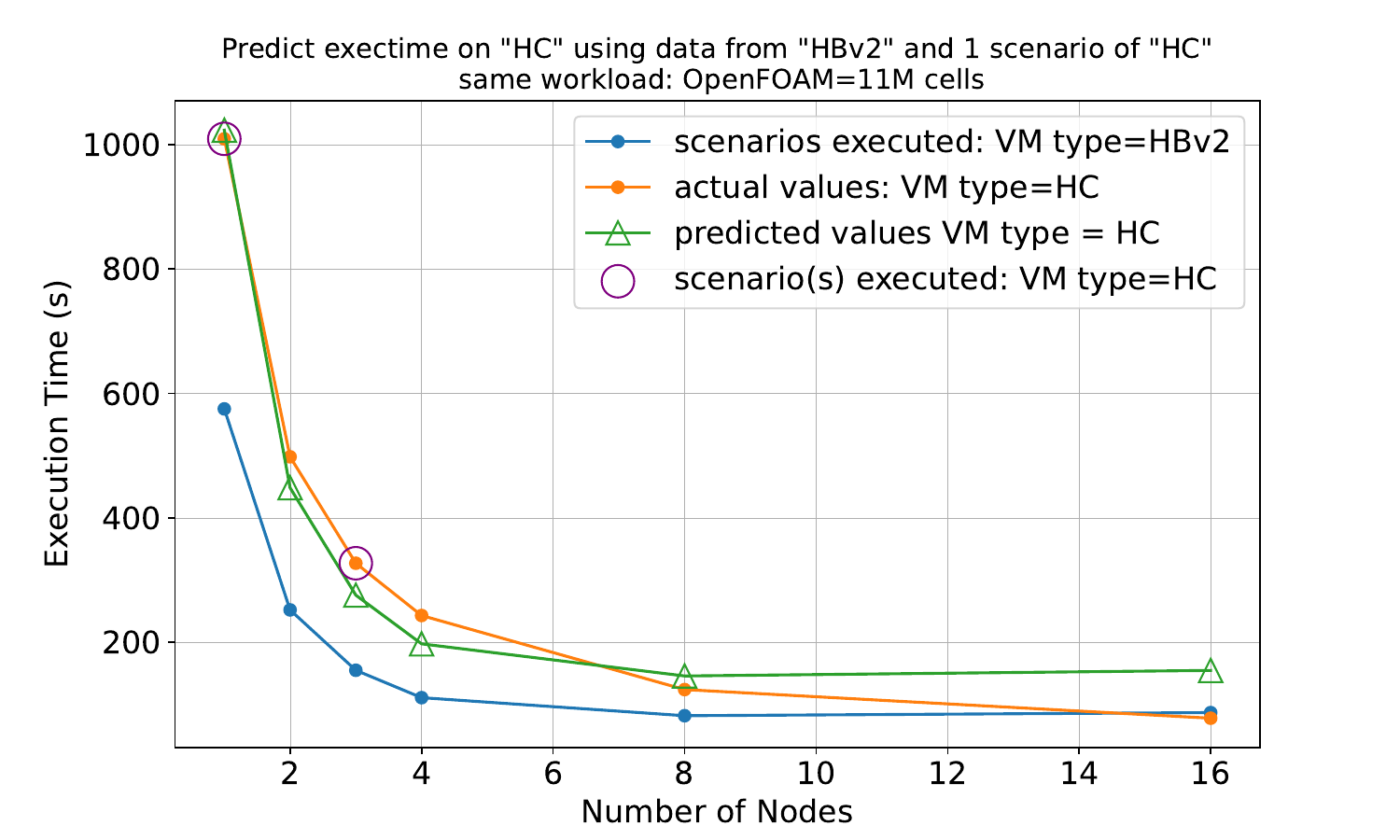}
       \vspace{-7mm}
        \caption{Predict the execution time curve using data from a different VM type and same application input---example for OpenFOAM.}
        \label{fig:fig1}
\end{figure}

\begin{figure}[!h]
        \centering
       \includegraphics[width=1.0\columnwidth]{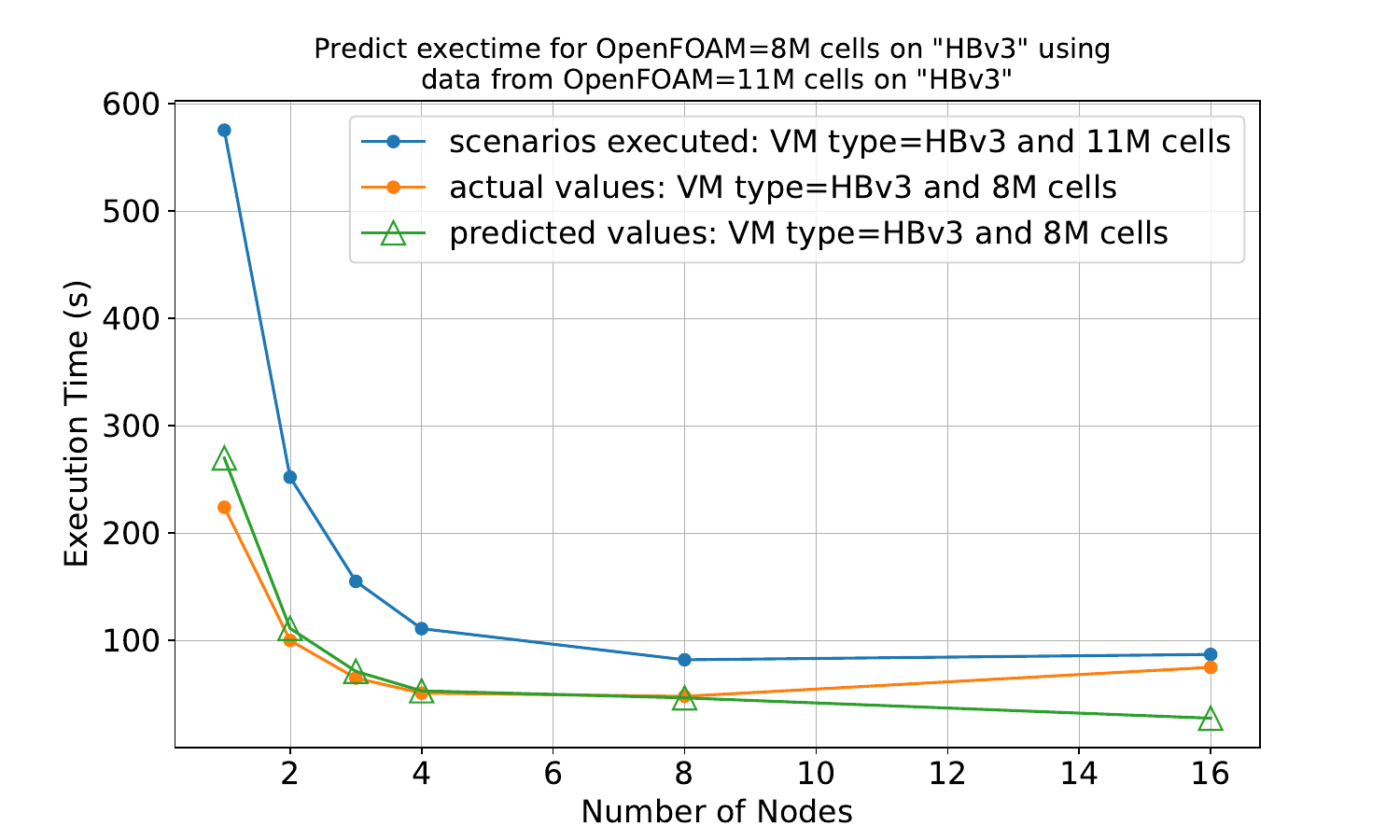}
       \vspace{-7mm}
        \caption{Predict the execution time for the same VM type but with different application input parameter---example for OpenFOAM.}
        \label{fig:fig2}
\end{figure}

\begin{figure}[!h]
        \centering
       \includegraphics[width=1.0\columnwidth]{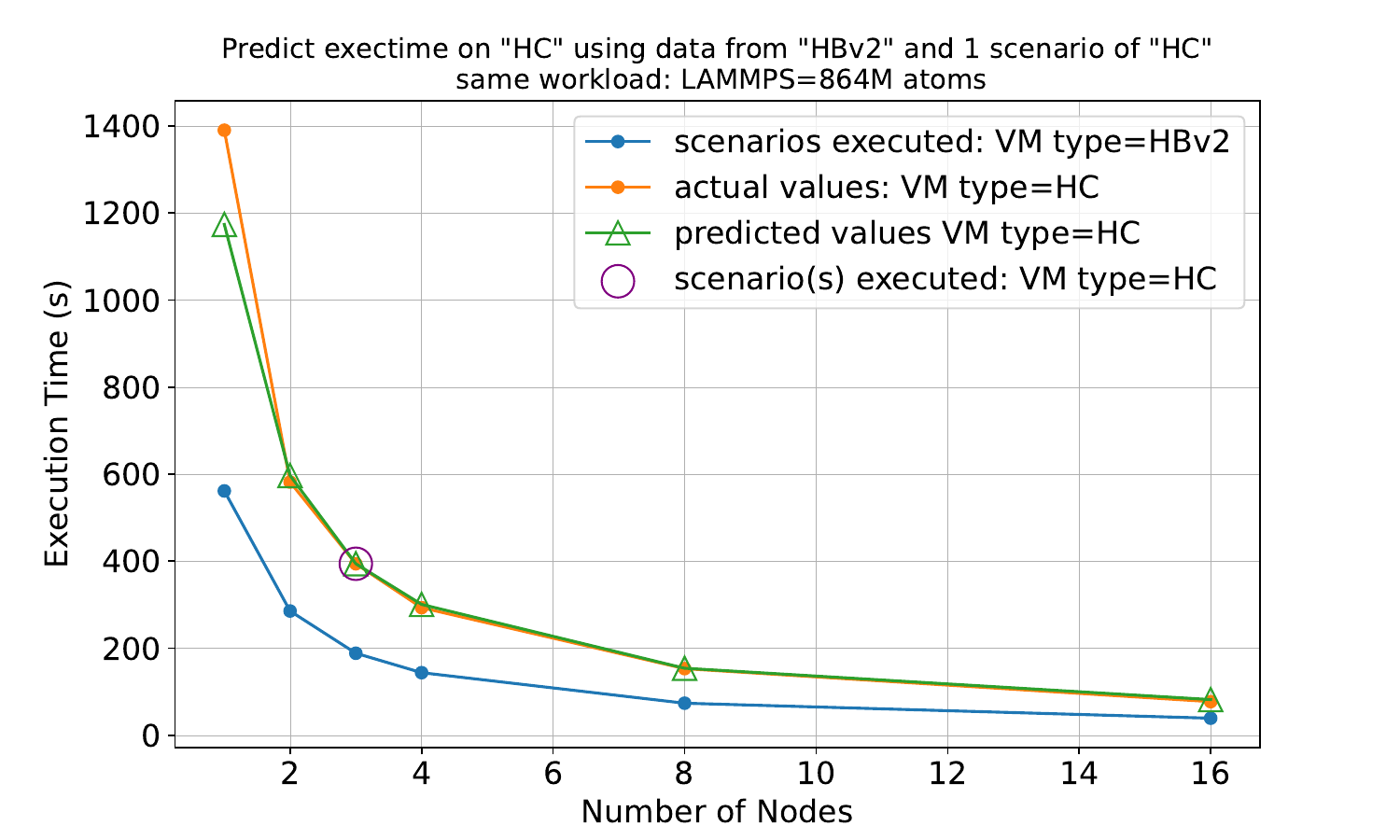}
       \vspace{-7mm}
        \caption{Predict the execution time curve using data from a different VM type and same application input---example for LAMMPS.}
        \label{fig:fig3}
\end{figure}

\begin{figure}[!h]
        \centering
       \includegraphics[width=1.0\columnwidth]{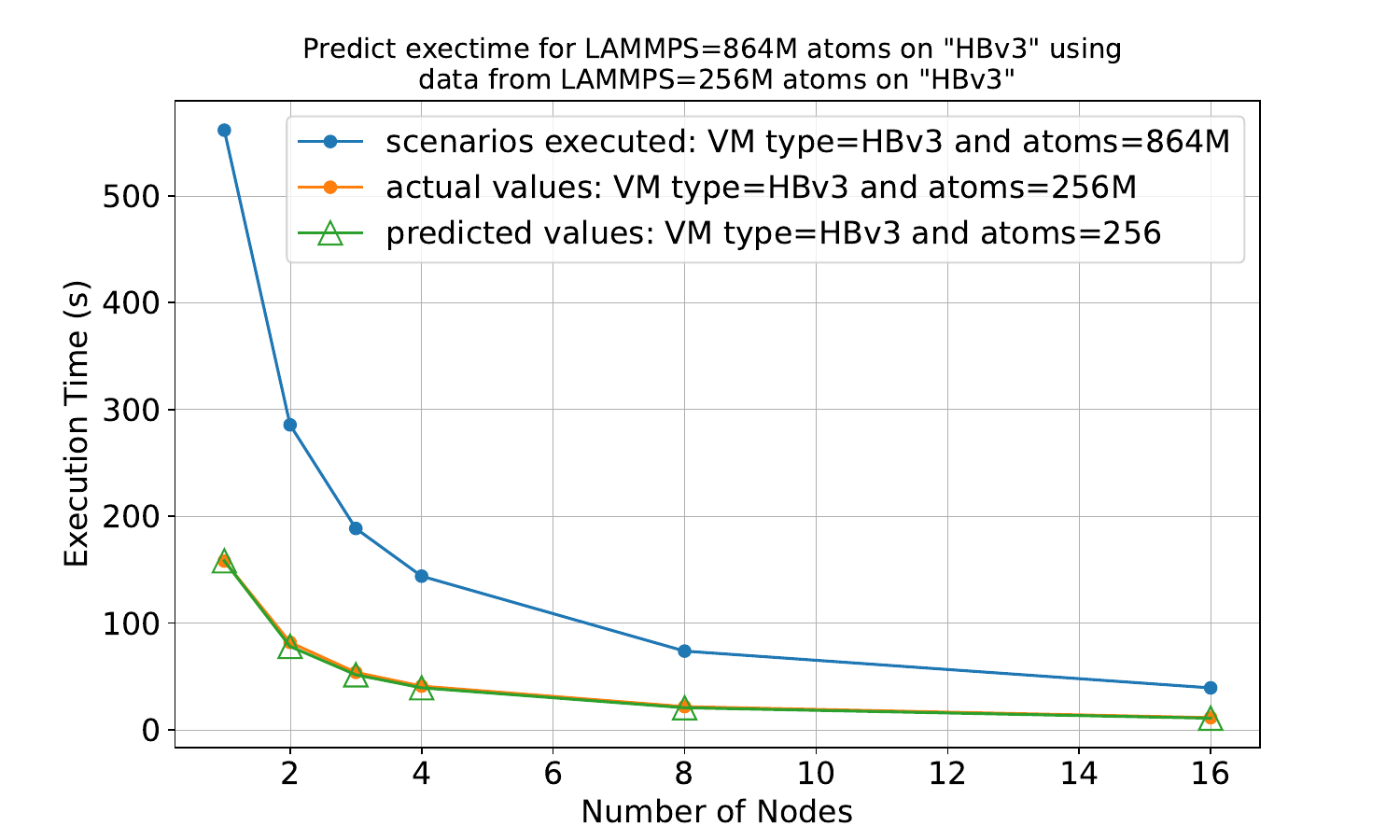}
       \vspace{-7mm}
        \caption{Predict the execution time for the same VM type but with different application input parameter---example for LAMMPS.}
        \label{fig:fig4}
\end{figure}

In the poster, we show three types of plots considering execution time, costs, and different numbers of VMs for two applications, OpenFOAM and LAMMPS (accessed via EESSI~\cite{droge2023eessi}), and three different values of an input parameter for each application. We also show predictions for the two cases above for both applications, illustrating that several scenarios can be eliminated when generating the recommendation of resource selection to users. For OpenFOAM we used the MotorBike benchmark and for LAMMPS we used Lennard-Jones benchmark.  We used three VM types: HC, HBv2, and HBv3, (AMD 44, 120, and 120 cores, respectively), interconnected by an InfiniBand network. And explored scenarios with up to 16 VMs, which is equivalent to 1,920 cores.

Our initial findings are that we can reduce a large percentage of scenarios to be executed in the cloud without requiring or generating extensive data (Figures~\ref{fig:fig1}, \ref{fig:fig2}, \ref{fig:fig3}, and~\ref{fig:fig4}). Some application input parameters have high influence on resource consumption, whereas others have none or minimal~\cite{lamar2023evaluating}. Thus, in practice, we can consider application input parameters for resource selection advice in the cloud, as long as there is a prior knowledge on their influence on application run times.

These findings are relevant to user workflows, such as, in OpenFOAM external aerodynamics, which typically involves running multiple simulations with varying geometries to meet performance targets.
Once a new project starts, new geometries and conditions require investigation. This often results in changes to mesh size or specific simulation parameters. HPCAdvisor can help identify the best cost/performance configuration using minimal data, like a single simulation, by leveraging historical simulation data.
As users adopt new hardware quickly, HPCAdvisor can speed up initial tuning, helping them quickly find the optimal setup for their calculations.

\section{Next Steps}

We will expand the experiments by considering more application input parameters and their influence on resource usage and execution time. We will also explore infrastructure metrics to reduce the number of required scenarios to generate the resource recommendations.

\bibliographystyle{IEEEtran}

\bibliography{references}

\end{document}